\documentclass[twocolumn,aps,prl]{revtex4-1}

\usepackage{amsmath}
\usepackage{graphicx}
\usepackage{esint}
\usepackage{setspace}
\usepackage{epsfig}
\usepackage{color}

\def\beq{\begin{equation}}
\def\eeq{\end{equation}}
\def\beqa{\begin{eqnarray}}
\def\eeqa{\end{eqnarray}}

\def\cH{{\mathcal H}}

\def\Tr{\mathrm{Tr}}

\newcommand{\bra}[1]{| #1 \rangle}
\newcommand{\ket}[1]{\langle #1 |}

\begin{document}  

\noindent\large{\textbf{}}\vspace{0.5cm}

\author{Jaroslav Vacek,$^{1}$ Jana Vacek Chocholou\v sov\'a,$^{1}$ Irena G. Star\'a,$^{1}$
Ivo Star\'y,$^{1}$  and
Yonatan Dubi $^{2}$}
\email{vacek@uochb.cas.cz;~jdubi@bgu.ac.il}
\affiliation{$^1$ Institute of Organic Chemistry and Biochemistry, v.v.i.,
Academy of Sciences of the Czech Republic, 
Flemingovo n\'am. 2, 16610 Prague 6, Czech Republic; E-mail: vacek@uochb.cas.cz~,$^2$Department of Chemistry and the Ilse-Katz Institute for Nanoscale Science and Technology, Ben-Gurion University of the Negev, Beer-Sheva 84105, Israel}

\title{Mechanical Tuning of Conductance and Thermopower in Helicene Molecular Junctions}

\begin{abstract}
Helicenes are inherently chiral polyaromatic molecules composed of all-ortho fused benzene rings possessing a spring-like structure. 
Here, using a combination of density functional theory and tight-binding calculations, it is demonstrated that controlling the length of the helicene molecule by mechanically stretching or compressing
 the molecular junction can dramatically change the electronic properties of the helicene, leading to a tunable switching behavior of the conductance and thermopower of the junction with on/off ratios of several orders of magnitude. 
Furthermore, control over the helicene length and number of rings is shown to lead to more than an order of magnitude increase in the thermopower and thermoelectric figure-of-merit over typical molecular junctions, presenting new 
possibilities of making efficient thermoelectric molecular devices. The physical origin of the strong dependence of the transport properties of the junction is 
investigated, and found to be related to a shift in the position of the molecular orbitals.\end{abstract}
\maketitle   






\section{Introduction}
Interest in single-molecule junctions arose from their originally proposed role as functional elements in electronic devices \cite{Aviram1974}. In recent years it has been realized that molecular junctions (MJs) can potentially fulfill additional functionalities ranging from opto-electronics and spintronics to thermoelectric 
energy conversion \cite{Aradhya2013,Tsutsui2012,Dubi2011}, which is the focus of this work. The thermopower and the thermoelectric figure of merit (FOM) are measures of a junction's ability to convert a temperature difference into electric power, and MJs were suggested as candidates for efficient and high-power thermoelectric devices \cite{Murphy2008,Bergfield2009,Bergfield2010,Finch2009,Karlstrom2011,Nozaki2010,Stadler2011,Nakamura2013,Garcia-Suarez2014,Evangeli2013} due to their low dimensionality, large versatility and low thermal  conductance \cite{Dubi2011,Mahan1996,Paulsson2003}. While there have already been several demonstrations of thermoelectric energy conversion in single-molecule junctions \cite{Malen2009,Malen2009a,Malen2010a,Reddy2007,Widawsky2012,Chang2014,Kim2014,Lee2014}, 
unfortunately the typical thermopower is rather small ($S\sim 5-50 ~\mu$V/K), much smaller than that of commercially available semi-conductor-based thermoelectrics. Increasing the 
thermopower of molecular junctions is an essential step if MJs are ever to become relevant energy conversion technologies. The natural question that arises in this context is then: 
is there a way to {\sl in-situ} increase the thermopower, $S$, and the thermoelectric FOM, $ZT$, of MJs?

\begin{figure}
\includegraphics[width=8.5truecm]{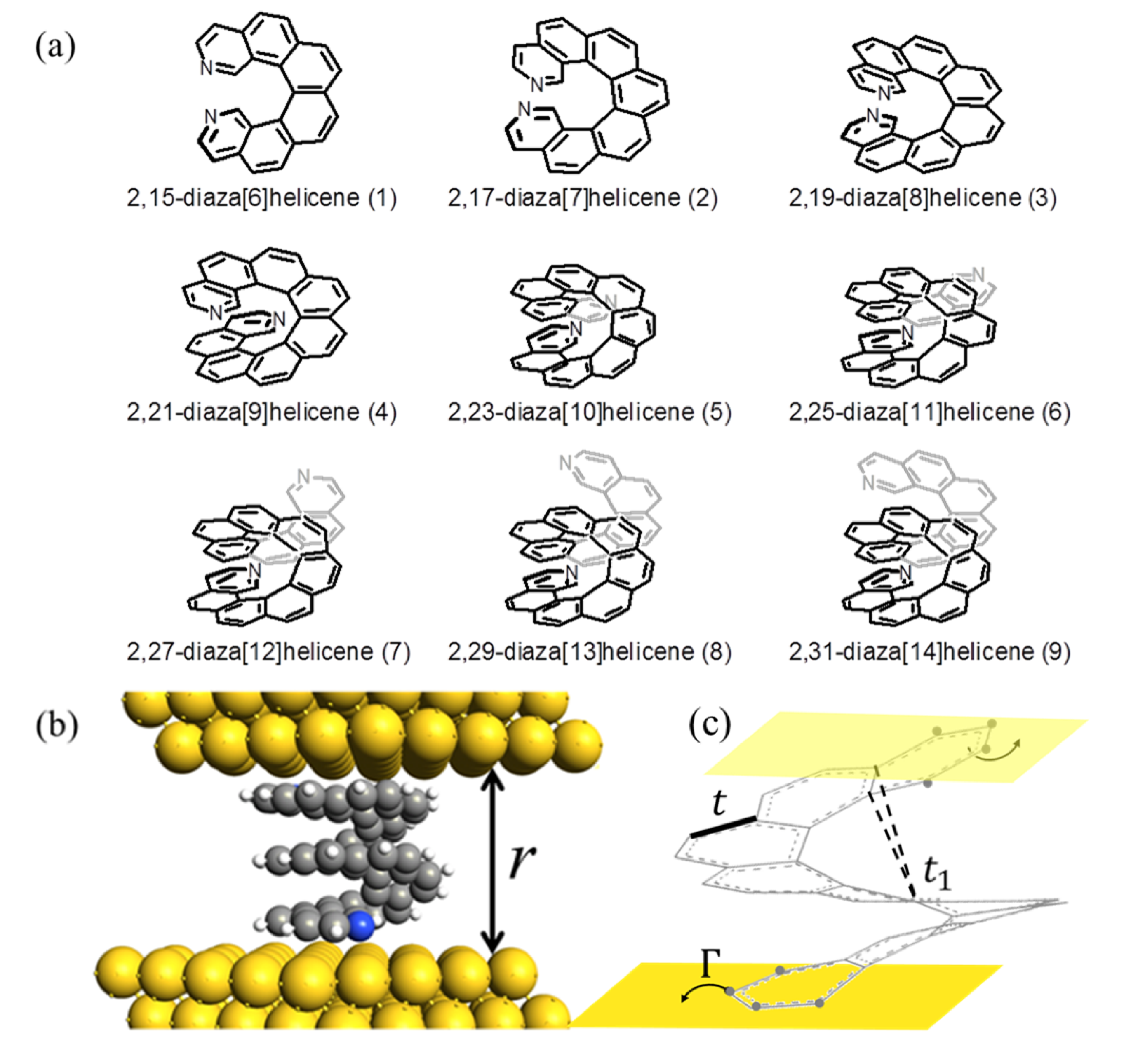}
\vskip 0.5truecm
\caption{Schematic illustration of the system studied: (a) Chemical structure of diaza[n]helicenes (1)-(9) used in the DFT and tight-binding calculations. Chirality issues were not a subject of this study, however, all the investigated systems were of (M) helicity. 
(b) 2,31-Diaza[14]helicene (9) molecule connected to two gold(1,1,1) contacts, $r$ denoting the distance between the electrodes. (c) Graphic representation of the tight-binding model of the helicene-based molecular junction. The tight-binding parameters for nearest neighbor interaction ($t$), inter-stack coupling ($t_1$) and molecule-electrode level broadening $\Gamma$ are noted. The four points in (c) mark the contact points of the rings with the electrode used for the tight-binding model.}
\label{fig1}
\end{figure}

In this letter, we propose to focus on helicene-based molecular junctions (HMJs), which can be used as a tool to help answer the above question. [n]Helicenes are 
aromatic compounds in which $n$ 
benzene rings or other (hetero)aromatics are angularly annulated to give helically-shaped molecules with a spring-like structure (Fig.~\ref{fig1}(a)). The chemistry of helicenes has been systematically 
explored for more than half a century (for recent reviews see, e.g. \cite{Shen2011,Gingras2013,Gingras2013a,Gingras2013b}), and a renewed and growing interest in helical aromatics is clearly visible 
within the last decade. Although the number of applications of helicenes is so far rather limited, they have already found astonishing applications to enantioselective organo- or transition metal 
catalysis \cite{Narcis2014,Yavari2014}, molecular recognition,\cite{Shinohara2010,Li2012} self-assembly, \cite{Shcherbina2009} nonlinear
 optical materials, \cite{Verbiest1998,Verbiest2001,Wigglesworth2005} chiral materials, \cite{Anger2012,Adriaenssens2011,Graule2009}  and chiroptical electronic devices. 
\cite{Yang2013,Fuchter2012,Shi2012} Furthermore, as a first step towards realizing single-molecule junctions, helicenes have been placed on surfaces and probed with atomic force and scanning tunneling 
microscopes. \cite{Stohr2011,Balandina2013,Prauzner-Bechcicki2010,Sehnal2009,Godlewski2012,Hauke2012,Fasel2006,Shchyrba2013}     
 
We suggest that the distance $r$ between the two metallic electrodes of a helicene molecular junction (Fig.~\ref{fig1}(b)) can be used as an experimental control parameter that can be tuned to alter 
and probe the properties of the HMJ. Due to the spring-like structure of the helicene molecule, when the junction is compressed, i.e., the distance between the electrodes of the junction is reduced, 
the couplings between the carbon orbitals on the benzene rings increases, opening additional transport paths 
for the electrons and coherently changing the electronic properties of the MJ. Regulation of the electrode 
distance in molecular junctions has already been introduced as a means to control and measure the molecular 
forces and couplings \cite{Zhou2013,Huang2007,Zhou2010,Diez-Perez2011} (interestingly, fullerene C60 was already demonstrated to minimize its resistance and 
become almost transparent for tunneling electrons if the molecule is sufficiently squeezed by an STM tip).
\cite{Joachim1995} However, in the studies mentioned above, the dominant origin for changes in transport is the change in the molecule-electrode coupling, while the molecular electronic structure remains largely unchanged. In a nonplanar molecule, on the other hand, pulling or stretching should affect both the molecular orbitals and the couplings. Nevertheless, there is a lack of suitable models of nonplanar molecules to 
systematically investigate the stretch- or compression-dependent single molecule conductance. 

We demonstrate our conceptual experiment by using density-functional theory (DFT)-based transport calculations to extract the transmission function 
of a HMJ for different values of the inter-electrode distance, $r$. To demonstrate how the helicene electronic 
structure depends on $r$, we then use a tight-binding (TB) model to fit the DFT data. Next, the TB model and parameters are used to predict the thermopower and $ZT$ of helicene molecular junctions as a function of $r$ and helicene length, 
demonstrating that these can be optimized to obtain maximal thermoelectric efficiency. It is important to point out early on that we are aiming to generate qualitative rather than 
quantitative predictions, and therefore the use of a TB approximation is justified. However, even when using the first-principle DFT calculation, the predictions can only be qualitative due 
to the lack of appropriate functionals and benchmark DFT calculations for helicenes. 

\section{Methods}
\subsection{Molecular system under study}
We studies diazahelicenes with the hetero-atoms in position 2 (symmetrically on both sides).  These systems were selected for three reasons:  (i) they are fully aromatic and highly conductive, (ii) 
they are synthetically accessible and were previously prepared by some of the authors \cite{Mivsek2008} and (iii) they provide for reasonably good binding to gold surfaces via the nitrogen functionalities. \cite{Bilic2002,Wu2006,Wu2003} For this reason, azamolecules such as 4,4'-bipyridine are often used in single molecule conductance experiments, for example break junction measurements. \cite{Quek2009}  Based on our DFT 
calculations, position 2 in helicenes appears more suitable for binding than, for example, position 1, for steric reasons. 

To achieve conductance and thermopower tuning and switching we used two additional parameters.  One is helicene length ([n]) which was varied between 6 and 14 aromatic rings and the other is 
helicene stretch/compression ($\Delta r$).  Out of the 9 total molecular systems of different number of rings, 2,15-diaza[6]helicene (1) to 2,31-diaza[14]helicene (9), only one was completely 
investigated by DFT methods.  
This was the 2,21-diaza[9]helicene (4) (Fig.~\ref{fig1}(a)) from the middle of the series.  Here we calculated all the stretched and compressed geometries, energies, and transmission spectra.  Due to 
the system size, the calculations were very lengthy and calculation of all 9 molecular structures would be impractical.  This was the motivation to use a simple tight-binding model to study the 
remaining systems at much lower computational cost.  The tight-binding model was, however, found to reproduce the DFT data surprisingly well. Chirality issues were not a subject of this study, all 
the investigated systems were of (M) helicity. 

\subsection{Density-functional-theory calculation}
\begin{figure}
\includegraphics[width=8.5truecm]{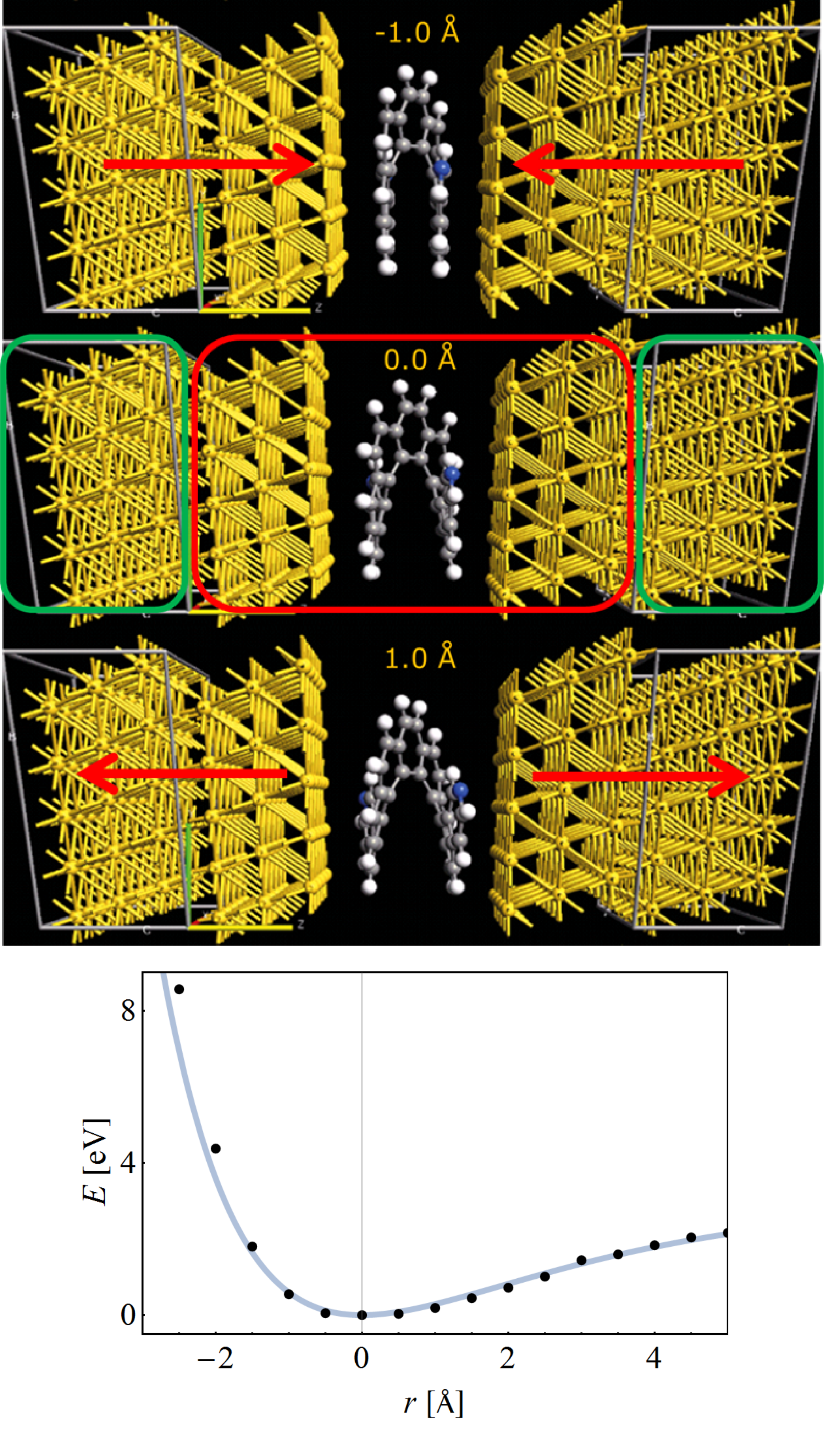}
\vskip 0.1truecm
\caption{Top panel: Two-probe model of the 2,21-diaza[9]helicene (4) HMJ used in the DFT geometry optimizations and transport calculations. Compressed (top), optimal (center) and stretched (bottom) structures. The green boxes in the optimal geometry indicate the left and right electrode regions, while the red box encircles the central region. Bottom panel: total energy $E$ as a function of $\Delta r$.}
\label{fig8}
\end{figure}

We considered a  2,21-diaza[9]-helicene(4) placed between two Au(111) electrodes. Prior to any periodic DFT calculations, all molecular structures were optimized in vacuum at the B97D/cc-pVDZ level of theory using the G09 package. 
\cite{Frisch}  QuantumWise Atomistix toolkit (ATK) \cite{Brandbyge2002,Soler2002,Stokbro2010} was then employed in all subsequent calculations described below.  First, the HMJ device was constructed as a "two-probe system" within 
the Virtual Nano Lab (VNL) module of ATK.  Here, the HMJ consists of three regions:  The left electrode (bulk Au), the so called central region and the right electrode (bulk Au), cf. Fig.~\ref{fig8} top panel.  In the central 
region the helicene molecule is sandwiched between six surface layers of Au(111), i.e. with three Au layers at both sides of the molecule.  The unit cell size of the left and right electrodes in the plane parallel to the gold 
surface was 5x5 Au atoms.  The entire two-probe system was then optimized and the electrode separation relaxed.  This resulted in our reference equilibrium ($\Delta r$ = 0, corresponding to an electrode separation of $\sim 8.5$ \AA) geometry from which we started the stretching and 
squeezing by just adjusting the electrode separation in 0.5 \AA~ steps.  The total compression attempted was 3.5 \AA~ while the stretching continued until the molecule broke off the electrodes at 7.0 \AA~ elongation. Only in the optimal configuration the unit cell was allowed to fully relax, so only limited deformations of the gold electrodes were allowed. The system 
was re-optimized at each step with the electrode separation constrained.  DFT-PBE/SZP level of theory with $2 \times 2\times 50$ k-point sampling was used for all optimizations while DFT-PBE/DZP with 3x3x100 k-point sampling was 
used for transmission spectra evaluation.  To check the numerical accuracy, optimizations with larger k-point sampling ($3\times 3\times 100$) were also tried with no difference in the resulting geometries. The bottom panel of Fig.~\ref{fig8} shows the total energy $E$ as a function of $\Delta r$. The forces required for stretching and squeezing can be estimated from the energy curve to be $\sim 1$ nN and $\sim 5$ nN, respectively. 

\subsection{Transport calculation within the tight-binding model}
The DFT results indicate that helicenes tend to lay flat on the electrode surface whenever possible \cite{Gingras2013,Seibel2014}.  At least one aromatic ring (but more than one for most geometries) tend to form contact with the 
electrodes for the electrode separations used in the TB and thermopower calculations, cf. Fig.~\ref{fig8}.  Of course, the contact weakens as we stretch the molecule out but a four-point contact model 
appeared to be applicable throughout the stretched/compressed structures used in this paper (Fig.~\ref{fig1}(c)). 

Once the Hamiltonian and self-energies are determined (Eq.~\ref{Hamiltonian}-\ref{Sigma}), all the transport properties can be determined within the Landauer formalism \cite{Datta1997,DiVentra2008,Peskin2010}. The Green's functions are determined via $G^{r,a}=\left( E-\cH_M+\Sigma^{r,a} \right) ^{-1}$, where $\Sigma^{r,a}=\Sigma^{r,a}_T
+\Sigma^{r,a}_B$, corresponding to the self-energies of the top and bottom electrodes. The transmission function is given by $T(E)=\Tr \left( \Sigma^r_T G^r \Sigma^a_B G^a \right) $, and the transport coefficients - the conductance $G$, the thermopower $S$, 
and the thermal conductance $\kappa$ - are determined within the Landauer formalism as $G=e^2 L_0,~S=L_1/(e T L_0,~\kappa=\left( L_2-\frac{L_1^2}{L_0}\right)/T$, where $T$ is the temperature (room temperature is considered hereafter) and 
$L_n=-\frac{1}{h}\int dE T(E)(E-\mu)^n \frac{\partial f}{\partial E}$ are the Landauer integrals, with $h$ being Planck's constant, $\mu$ the chemical potential of the electrodes (set as the 
zero of energy hereafter), and $f(E)$ the Fermi-Dirac distributions. The thermoelectric figure of merit $ZT$ is given by $ZT=\frac{G S^2}{\kappa/T}$.

\section{Results: tuning conductance and thermopower }
\begin{figure}[h!] 

\includegraphics[width=9truecm]{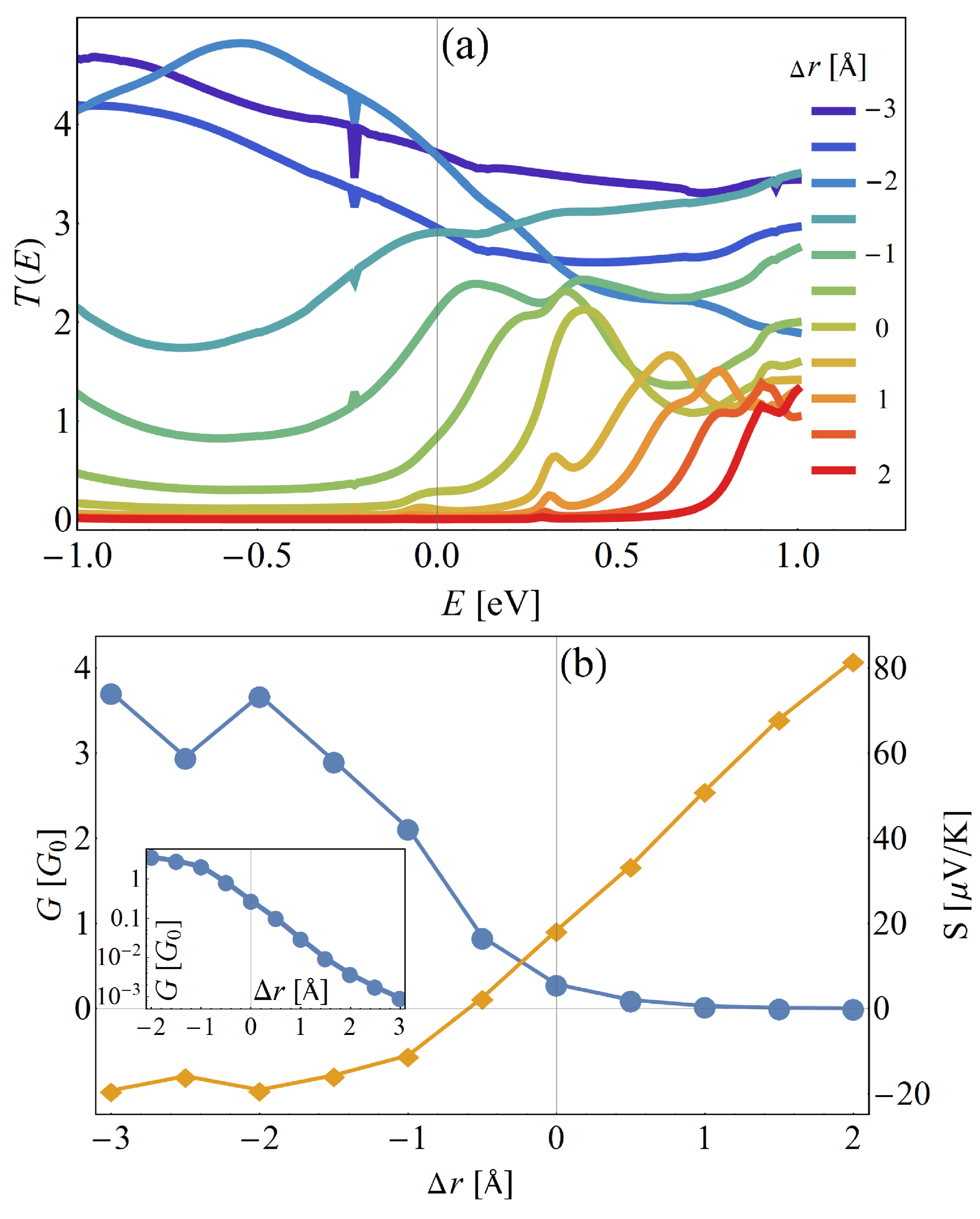}

\caption{Transmission and thermopower properties of HMJ formed by 2,21-diaza[9]helicene (4) and gold(1,1,1) electrodes:(a) Transmission $T(E)$ as a function of energy $E$ for different inter-electrode distances incrementally 
changed about $\Delta r=-3,2.5,-2,...,1,2$\AA ~(measured from the relaxed inter-electrode distance). (b) Conductance (blue circles) and thermopower (yellow diamonds) as a function of $ \Delta r$. Inset: Conductance on a 
logarithmic scale, demonstrating the order of magnitude change in conductance with changing $\Delta r$. }
\label{fig2}
 \end{figure}

The starting point is thus a DFT-based calculation of the transmission through a helicene molecular junction, formed by 2,21-diaza[9]helicene (4) and two gold(1,1,1) electrodes. Chirality issues were not a subject of this study, however, all the investigated systems were of (M) helicity. The transmission curves were calculated with the combined DFT-non-equilibrium Green's function approach \cite{Brandbyge2002,Soler2002,Stokbro2010}. Fig.~\ref{fig2}(a) shows the 
transmission curves $T(E)$ as a function of energy $E$ for different inter-electrode distances $r$, incrementally changed about $\Delta r=-3,-2.5,-2,...,1,1.5$\AA ~, measured from the relaxed inter-electrode distance $r$ (it is important to point here that $\Delta r=0$ 
is the distance where the DFT calculation finds the energy minimum, but may not correspond to any experimental distance, since in any experimental setting the distance will be controlled by an external force). The most 
striking feature of Fig.~\ref{fig2}(a) is the apparent closing of the transmission gap between the highest occupied and lowest unoccupied molecular orbitals (HOMO and LUMO, respectively) upon compression of the molecular junction, 
which transforms the MJs electronic nature from "insulating" to "metallic", by which we mean resonant tunneling (close to a molecular level).  Consequently, the conductance changes by more than three orders of magnitude. In Fig.~
\ref{fig2}(b) the conductance (in units of the conductance quantum $G_0=2e^2/h$) and the thermopower are plotted as a function of the distance $r$. The conductance shows a switching behavior with an "on"/"off" ratio of more than 
three orders of magnitude (inset of Fig.\ref{fig2}(b)), suggesting that HMJs can perform as reversible, nanoscale molecular switches. \cite{Cai2005,Kronemeijer2008,Lau2004,Molen2010,Loh2012,Diez-Perez2011,LiHuang2012,Martin2009,Martin2011,Kim2011} The 
thermopower shows a sign-change, a direct experimental prediction. this is an important feature, since a sign-change of the thermopower implies a change in the transport mechanism of the junction \cite{Dubi2011}, yet such a change 
is hard to demonstrate {\sl in situ}. 

To understand the origin of this change in the helicene electronic structure, we construct a TB description of the system. The use of TB models to complement DFT calculations is rather common (see, 
e.g. \cite{Dubi2013a,Karlstrom2014,Viljas2008,pauly2008}), and is useful for several reasons: (i) DFT calculations may become extremely computationally expensive and lengthy, (ii) the TB model allows for 
interpretation of the DFT calculation in terms of a simple physical picture, which is in many cases hard to build from the output of the DFT calculation (typically energies, orbitals and transmission), (iii) additional effects (such as dephasing, see below) cannot be introduced within the DFT calculation. The helicene molecule TB Hamiltonian is given by
\beq
\cH_M=\sum_i \epsilon_0 \bra{i}\ket{i}-t\sum_{\langle i,i' \rangle} \bra{i}\ket{i'}-t_1 \sum_{\{ i,j \}} \bra{i}\ket{j} ~~, \label{Hamiltonian}
\eeq
where $\bra{i}$ represents the orbital located on the $i$-th atom in the helicene, $\epsilon_0$ is the orbital energy, $t$ is the nearest-neighbor hopping matrix elements (with 
$\langle i,i' 
\rangle$ representing nearest-neighbor orbitals), and $t_1$ represents hopping matrix elements between sites that reside vertically one above the other on adjacent turns of the helix, so-called inter-stack hopping (with $\{ 
i,j \}$ representing vertically-neighboring orbitals). $t$ and $t_1$ are graphically depicted in Fig.~\ref{fig1}(c). The electrodes are treated within a wide-band approximation \cite{Verzijl2013} through a self-energy term, 
\beq
\Sigma^{r,a}_{T,B}=\mp i \Gamma_{T,B} \sum_{i_{T,B}} \bra{i_{T,B}}\ket{i_{T,B}}~~,
\label{Sigma}
\eeq
where $r(a)$ stands for retarded (advanced) self energy, $T$ and $B$ stand for top- and bottom-electrodes, respectively, and $\Gamma_{T,B}$ are the corresponding electrode-induced 
level-broadenings (we set $\Gamma_T=\Gamma_B=\Gamma$ for simplicity hereafter). The indices $i_{T,B}$ represent the sites on the helicene molecule that are in contact with the electrode 
(see Fig.~\ref{fig1}(c)). We assume the edge benzene rings of the helicene couple with the electrodes via the four external orbitals (that is, the edge rings lie almost flat on the 
electrodes). We found that changing the contact configuration does not qualitatively change our results. 

\begin{figure}[t]
\vskip 0.2truecm
\includegraphics[width=8.5truecm]{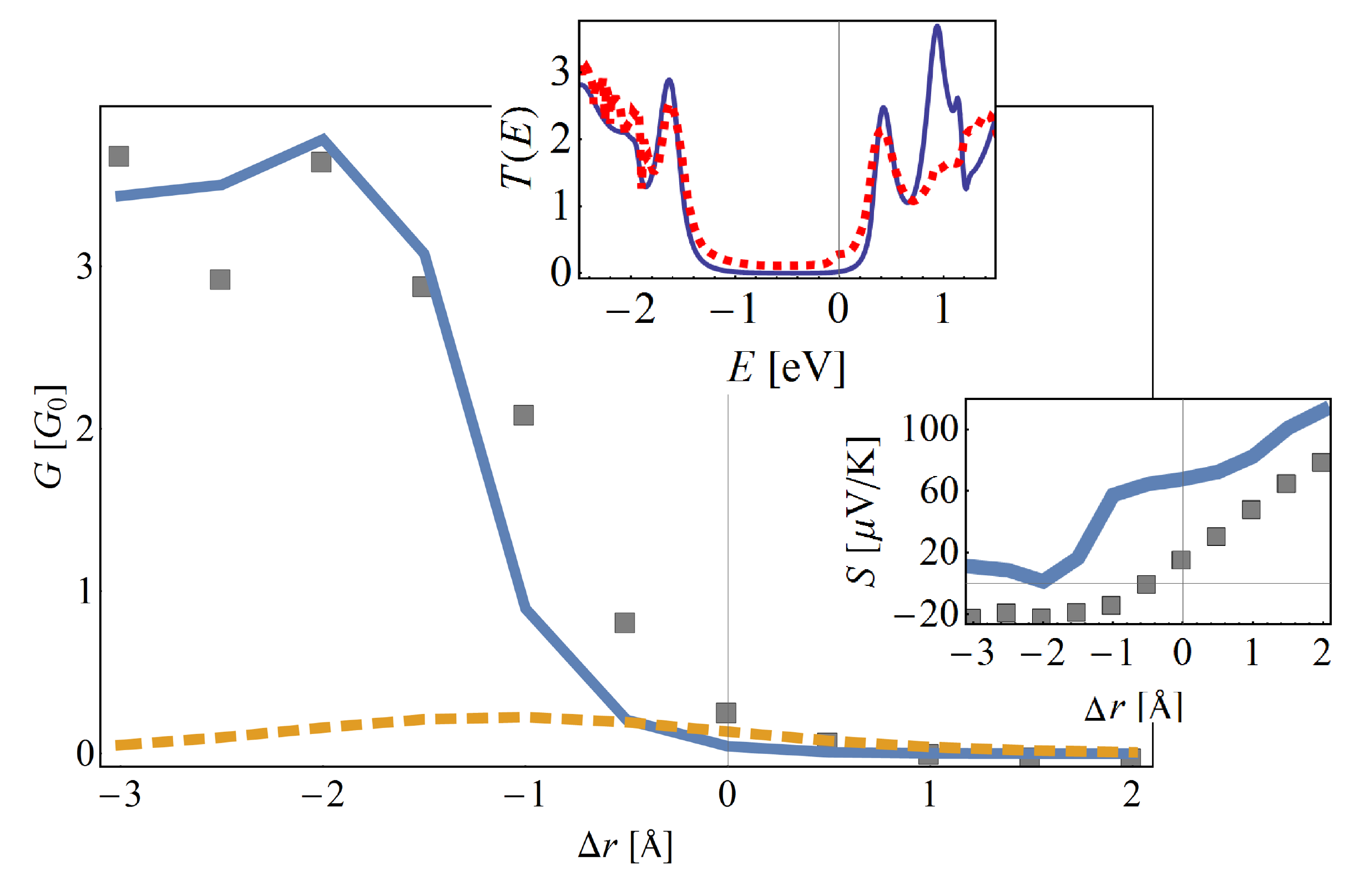}

\caption{2,21-Diaza[9]helicene (4): Conductance at the Fermi energy level as a function of the change $\Delta r$ in the inter-electrode distance measured from the relaxed configuration, based on the DFT data (filled gray squares) and on the TB calculation, assuming that both $t_1(r)$ and $\Gamma(r)$ depend exponentially on $r$ (solid blue line). The dashed orange line shows the same for a TB calculation where $t_1$ is independent of $r$ (see text for parameter values).
Top inset: the transmission as a function of energy for the DFT data (solid blue line) and the TB calculation (dashed red line). Bottom inset: Thermopower $S$ as a function of $\Delta r$ based on the DFT data (filled gray squares) and TB calculation (solid blue line). }
\label{fig3}
\end{figure}

To obtain the TB parameters, parameters $\epsilon_0,t,t_1$ and $\Gamma$ are tuned for a best fit between the transmission functions as obtained from the tight-binding 
and from the DFT for a HMJ (formed by 2,21-diaza[9]helicene (4) and gold (1,1,1) electrodes) at $\Delta r=0$. In the top inset of Fig.~\ref{fig3} the transmission curves of the DFT (solid blue line) and TB (dashed red line) calculations are shown, with the best 
fit yielding the tight-binding parameters $\epsilon_0=-0.61$ eV, $t=1.98$ eV, $t_1=0.279$ eV and $\Gamma=1.78$ eV. The TB curve fits the DFT data very well, even with such a small number of 
parameters and especially around the HOMO and LUMO resonances and below the HOMO level (note that for transport calculations, the important fit region is within $k_B T$ from the Fermi level).

To capture the HMJ electronic behavior under compression or stretching, it is necessary to know the $r$-dependence of the TB parameters. It is safe to assume that the orbital energy $\epsilon_0$ 
and the nearest-neighbor hopping element $t$ do not change substantially when the helicene molecule is stretched, and that most of the change occurs in $t_1$ and $\Gamma$. Based on knowledge that the 
dependence of $t_1$ and $\Gamma$ on $r$ originates from changes in the overlap of wave-functions, a reasonable assumption is that they depend exponentially on $r$, 
$t_1(r)=t_1 \exp \left(-\frac{r}{\xi_{t}}\right),~\Gamma(r)=\Gamma  \exp \left(-\frac{r}{\xi_{\Gamma}}\right)$. The values of $\xi_t$ and $\xi_\Gamma$ are found by fitting the TB transmission at 
the Fermi energy $T(\epsilon_F)$   (which at low temperatures is proportional to the conductance \cite{DiVentra2008,Peskin2010}) to the DFT data. In Fig.~\ref{fig3}, the conductance is plotted as a 
function of the the change in the inter-electrode distance, $\Delta r$, obtained from the DFT data (filled gray squares) and the TB calculation (solid blue line). This fit yields the values $\xi_t=1.33$\AA~ and $\xi_\Gamma=1.2$
\AA~. To demonstrate that it is necessary to include the change in $t_1$, the dashed orange line of Fig.~\ref{fig3} shows $T(\epsilon_F)$  as a function of $r$ for the case where only $\Gamma$ is 
dependent on $r$ and $t_1$ is independent of $r$ (i.e. where the transport dependence on $r$ comes only from the change in the contacts, while the electronic structure of the molecule itself is 
unchanged). The striking difference between the dashed orange line and the DFT calculation (filled gray squares) clearly shows that the internal electronic structure of the molecule is changed upon 
stretching (via a change in $t_1$). The bottom inset of Fig.~\ref{fig3} shows the thermopower $S$ as a function of $r$ based on the DFT data and TB calculation (solid blue line). While some shift is 
visible, the overall order of magnitude and trend are similar, which is impressive considering the small number of parameters in the model and the
 fact that thermopower is extremely sensitive to small variations in the parameters of the molecular junction \cite{Dubi2013a,Dubi2013b} (note that the thermopower is approximately proportional to the 
logarithmic derivative of the transmission function).

\begin{figure}
\includegraphics[width=8.5truecm]{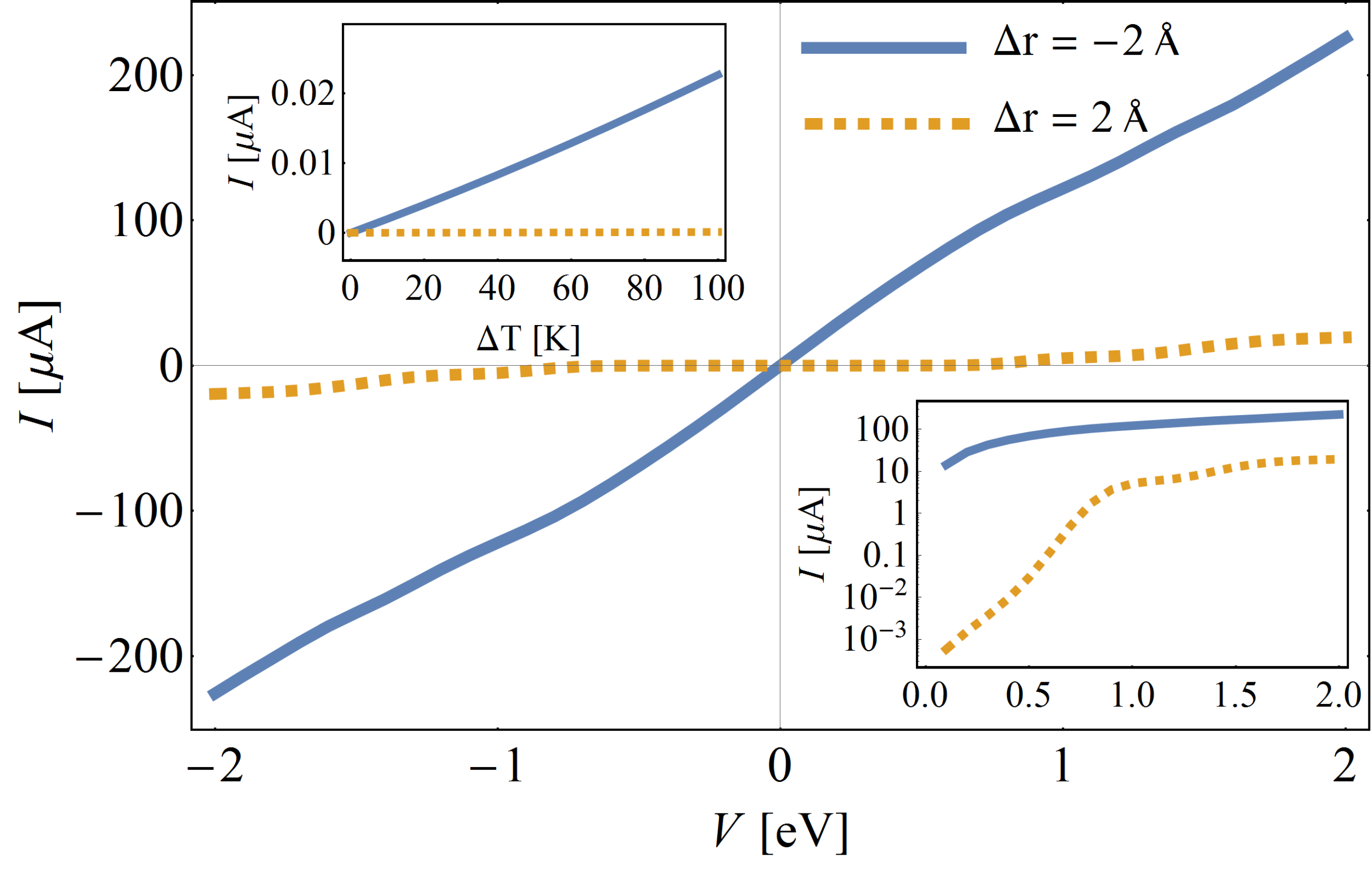}
\vskip 0.1truecm
\caption{Main panel:  $I-V$ curve of a 2,21-diaza[9]helicene (4) HMJ for two values of stretching distance, $\Delta r=-2$\AA~  (solid blue line) and $\Delta r=2$\AA~  (dashed yellow line). A switching ratio of $\sim 4$ orders of magnitude between the "On" and "Off" state can be seen in the bottom-right inset, where the same is plotted on a log scale. Top-left inset: the thermo-current $I$ as a function of temperature difference $\Delta T$ for the two states, showing a switching behavior.}
\label{fig4}
\end{figure}

To demonstrate the potential of HMJs as switches, in Fig. ~\ref{fig4} the $I-V$ curve of a [9]helicene HMJ is plotted for two values of stretching distance, $r=-2$\AA~ (solid blue line) and $r=2$\AA~ (dashed yellow line). The prominent dissimilarity  between them shows that the HMJ can be mechanically switched from a "metallic" 
("on" state) to an insulating ("off") state. The bottom-right inset shows the same on a log scale, to more clearly show the four orders of magnitude change in the current for low biases. 

Besides an electronic switch, HMJs can serve as a {\sl thermoelectric} mechanical switch, i.e. a junction who's thermoelectric current - the current due to a temperature 
difference - can change substantially. In the top-right inset of Fig.~\ref{fig4}, the current is plotted as a function of temperature difference $\Delta T$ (at zero bias) for $r=-2$\AA~ (solid blue 
line) and $r=2$ \AA~ (dashed yellow line). We set the temperature of the right electrode to be room temperature, $T_R=300$K, and the temperature of the left electrode is $T_L=300+\Delta T$. The 
thermo-current shows $\sim 4$ orders of magnitude difference between the two states at $\Delta T \sim 50$K (in fact, since the thermopower changes sign, in principle one can have perfect 
thermo-electric switching because the "off" state can be tuned to have zero thermo-current). 

The dramatic change of conductance upon mechanical stretching (or squeezing) is a direct prediction that is accessible to state-of-the-art experiments. In addition, since the whole effect is based on 
a coherent calculation of transport, it is of interest to examine the role of dephasing and decoherence in this system. Indeed, the majority of the theoretical approaches to transport and thermopower 
in MJs implicitly assume that the transport is coherent, and that there is no dissipation within the molecular junction itself, only at the electrodes or at the molecule-electrode interface 
\cite{Datta1997,DiVentra2008,Scheer2010}. Recent experiments provide some evidence for coherent transport in the form of a zero-bias conductance dip 
\cite{Aradhya2012,Arroyo2013,Fracasso2011,Guedon2012,Hong2011,Rabache2013,Batra2014}. However, these experiments typically compare between different junctions (measured in separate experiments), and 
lack a control parameter that can be tuned to verify the presence of coherence {\sl in situ} (except temperature in some cases \cite{Markussen2014}). As we demonstrate below, the dependence of  
conductance on electrode distance $r$ can be an additional experiment to show coherent (or incoherent) transport in MJs. 

\begin{figure}
\includegraphics[width=8.5truecm]{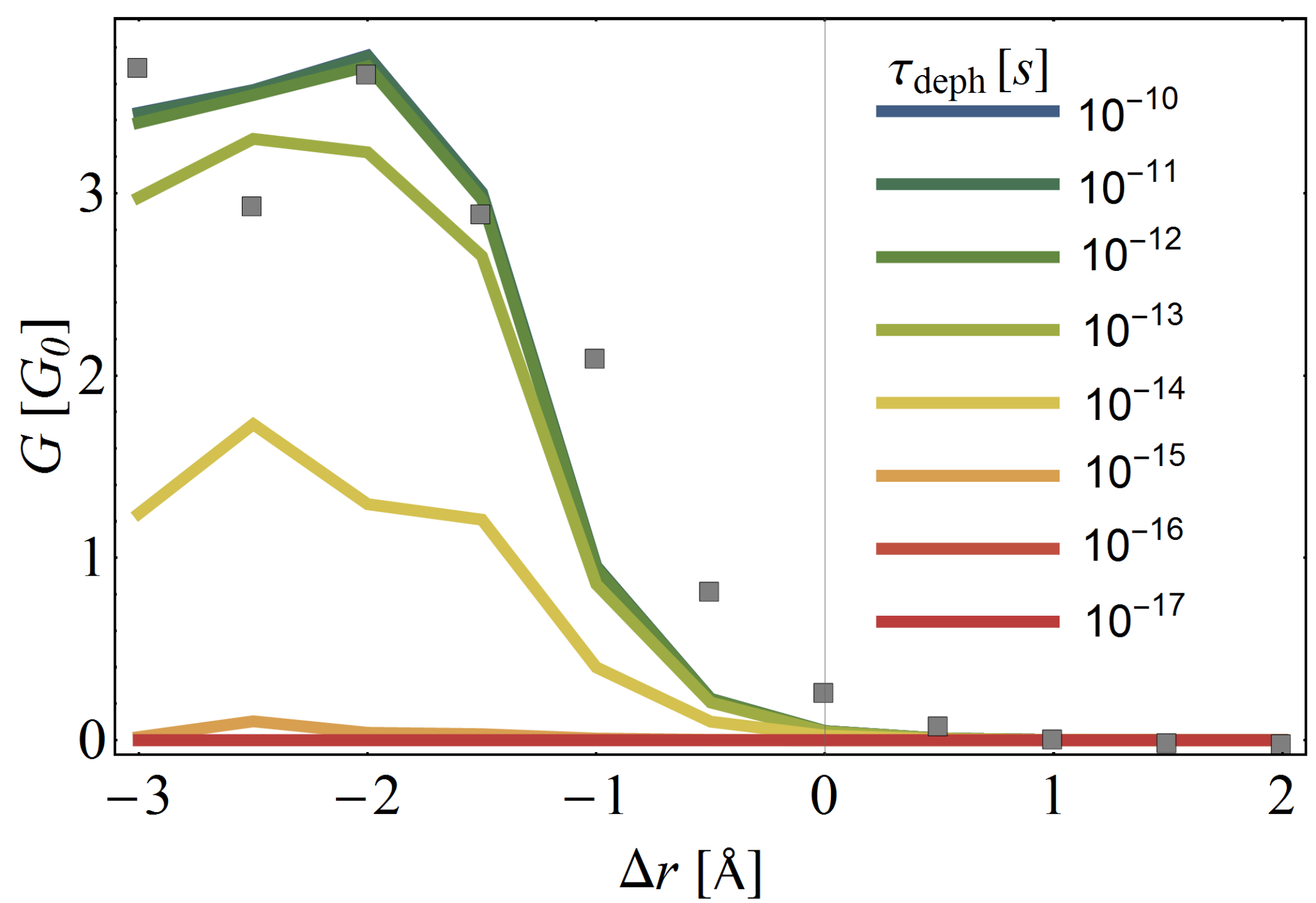}
\vskip 0.1truecm
\caption{Conductance as a function of the change $\Delta r$ in the inter-electrode distance measured from the relaxed configuration, based on the DFT data (filled gray squares) and on the TB 
calculation, in the presence of dephasing, for different values of the dephasing time, $\tau_{\mathrm{deph}}=10^{-10},10^{-11},...,10^{-17}$s. }
\label{fig5}
\end{figure}

For this aim, we repeat the calculation of conductance vs. $\Delta r$ (as in Fig.~\ref{fig3}), now in the presence of dephasing. We treat dephasing in a phenomenological way, by adding to the 
self-energy (Eq.~\ref{Sigma}) a diagonal term $\Sigma^{(r,a)}_{\mathrm{deph}}=-i \Gamma_{\mathrm{deph}}$, where $\Gamma_{\mathrm{deph}}=\hbar / \tau_{\mathrm{deph}}$ and $\tau_{\mathrm{deph}}$ is the 
dephasing time, i.e. the average time it takes for an electron residing on the helicene molecule to loose its phase \cite{DAmato1990,Krems2009,Nozaki2012,Pal2011,DiVentra2008}. In Fig.~\ref{fig5}, the 
conductance as a function of $\Delta r$ is plotted for different values of the dephasing time, $\tau_{\mathrm{deph}}=10^{-10},10^{-11},...,10^{-17}$s (as in Fig.~\ref{fig3}, the squares are the data from the 
DFT calculation, which is purely coherent). As seen, dephasing has a substantial effect on the conductance tunability, and dephasing times of $\sim 10^{-16}-10^{-17}$ are in fact detrimental to the 
switching behavior. 

\section{Results: thermoelectric figure of merit}
\begin{figure}
\includegraphics[width=8.5truecm]{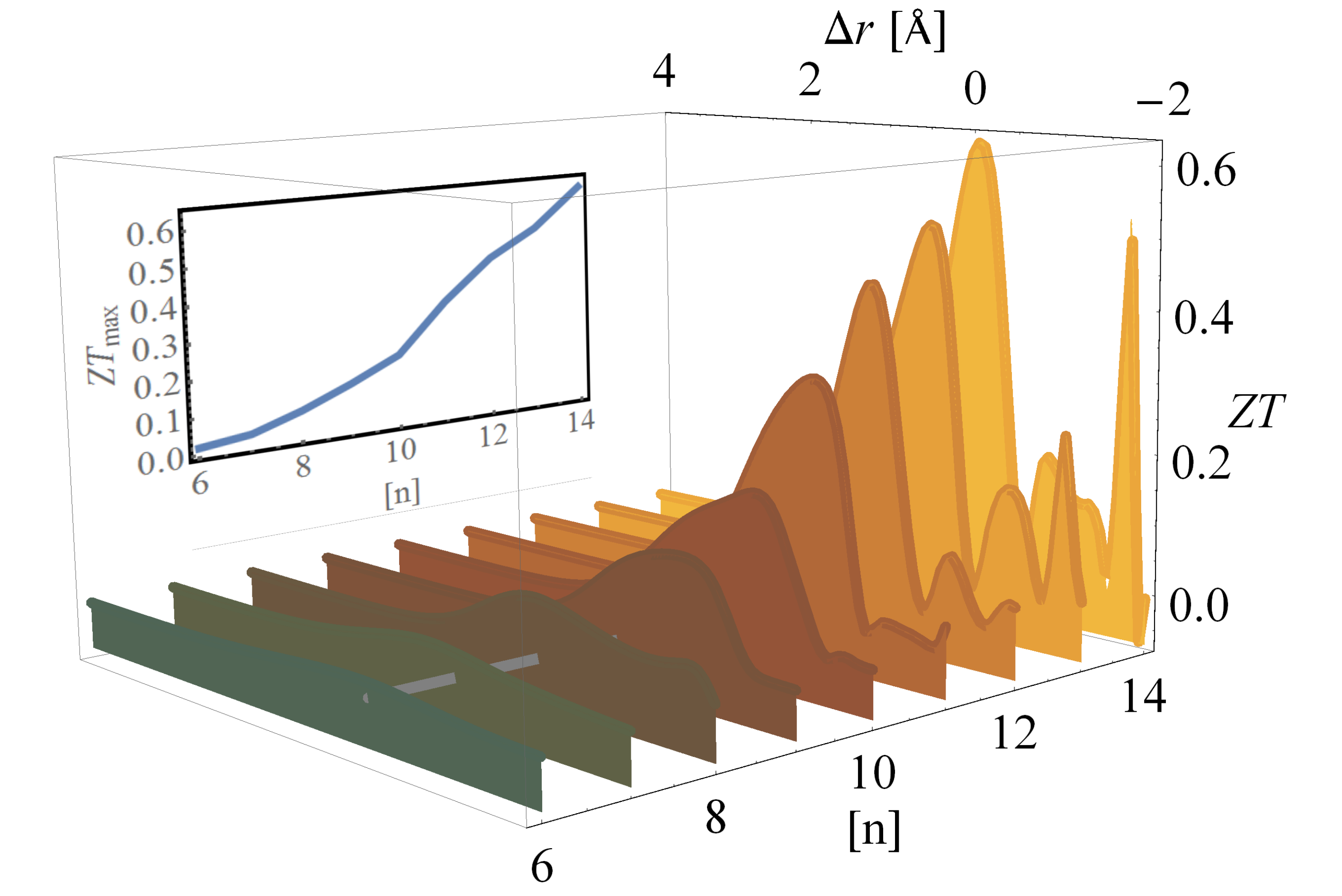}
\vskip 0.1truecm
\caption{$ZT$ as a function of $\Delta r$ for a series of homologous diaza[n]helicenes (from 2,15-diaza[6]helicene (1) to 2,31-diaza[14]helicene (9)). Inset: maximal thermoelectric FOM, $ZT_{\mathrm max}$, as a function of helicene length [n] (number of fused (hetero)aromatic rings). }
\label{fig6}
\end{figure}

With the TB parameters (and their dependence on inter-electrode distance $r$) determined, we can proceed to evaluate the thermoelectric FOM of HMJs. In Fig.~\ref{fig6},  $ZT$ is 
plotted as a function of $\Delta r$ for diaza[n]helicenes (1)-(9) differing in the number of all-ortho fused (hetero)aromatic rings (n = 6, 7, ... 14) and, accordingly, in their length. To calculate $ZT$ realistically, we take also the phonon thermal conductance into account, with a typical value of $\kappa=20 {\mathrm pW}/{\mathrm K}$ \cite{Malen2010,Ong2014,Segal2003,Nozaki2010}.  Two conclusions can be quickly drawn from this figure: (i) $ZT$ can be tuned by changing $r$, and displays order of 
magnitude difference within the same junction, and (ii) the longer the helicene the better its thermoelectric performance. We find that the optimal $ZT$ is obtained at $\Delta r\sim 0$. In the inset the 
maximal $ZT$, $ZT_\mathrm{max}$ is plotted as a function of helicene length (solid line), showing that $ZT_\mathrm{max}$  values can be as large as $ZT_\mathrm{max}\approx 0.6$, which far exceeds current MJ values. 

\begin{figure}
\includegraphics[width=8.5truecm]{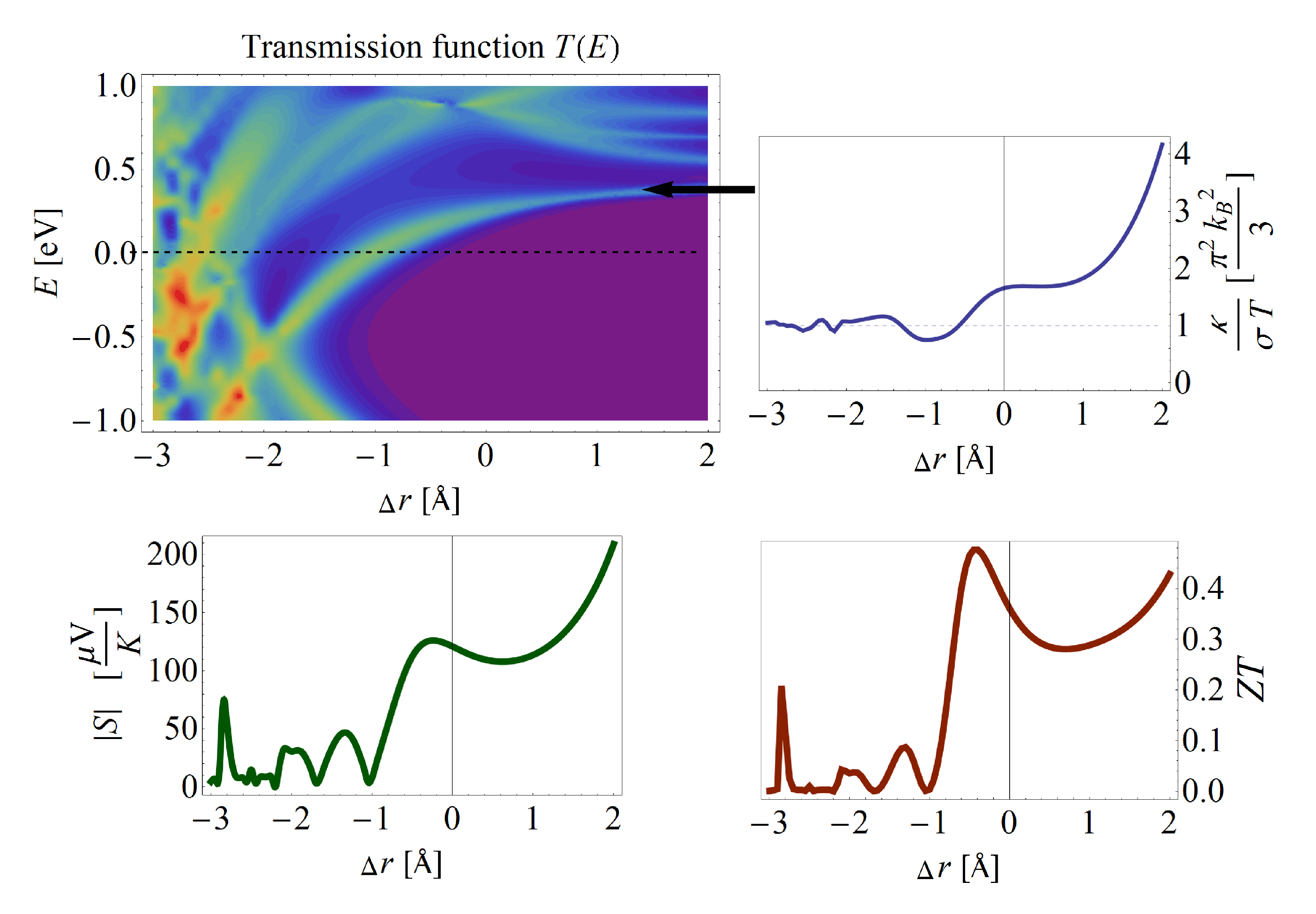}
\vskip 0.1truecm
\caption{(a) Color-plot of the Transmission as a function of energy $E$ and inter-electrode distance $\Delta r$. (b) Thermopower (absolute value) $|S|$ as a function of $\Delta r$. (c) The Lorentz number 
 $\kappa /\sigma T$ as a function of $\Delta r$. (d) $ZT$ as a function of $\Delta r$. The calculation is done for 2,27-diaza[12]helicene (7) (n = 12, see text for other numerical parameters).}
\label{fig7}
\end{figure}

To understand the origin of this strong $ZT$ dependence on $r$ we focus on the 2,27-diaza[12]helicene (7) HMJ ([n] = 12) HMJ as an example (but the results are qualitatively similar for all [n]), and 
study how the different elements that 
constitute $ZT$ depend on $\Delta r$. In Fig.~\ref{fig7}(a) the transmission function $T(E)$ is color-plotted as a function of $E$ and $\Delta r$. The black arrow indicates the position of the LUMO peak, and it can 
be seen that upon compression of the junction (which increases the inter-stack hopping term $t_1$), the position of the LUMO changes, shifting down toward the Fermi level (dashed line). 
Once the LUMO crosses the Fermi, the HOMO-LUMO gap is no longer visible and the HMJ becomes "metallic" (resonant tunneling). The shift in the LUMO position implies that the filling of the gap cannot 
be viewed as due to the emergence of an addition conduction channel (the $\pi$-stacking), but rather due to a change in the electronic properties of the helicene, induced by enhanced hopping along  
the stacks.

The thermopower $S$, whose absolute value is plotted in Fig.~\ref{fig7}(b), reflects the behavior of $T(E)$ (since it is 
approximately equal to the logarithmic derivative of $T(E)$ at the Fermi level \cite{DiVentra2008,Dubi2011}). The peak in $|S|$ at $\Delta r\sim -0.5$\AA ~ is a result of the Fermi level being close to (but not at) the LUMO 
resonance, where the slope of $T(E)$ is maximal. Surprisingly, $|S|$ increases dramatically when the junction is stretched, although the LUMO resonance shifts away from the Fermi 
level. This is due to the decrease in $\Gamma$, which typically increases the slope of the transmission function. Note that the thermopower can reach values of $S\sim 100-200 \mu {\mathrm V}/{K}$ (and larger for longer HMJs), which is more than one order of magnitude larger than values typically observed in molecular junctions \cite{Malen2009,Malen2009a,Malen2010a,Reddy2007,Widawsky2012}. 

In Fig.~\ref{fig7}(c) the ratio between the thermal and charge conductances $\kappa /\sigma T$ is plotted as a function of $\Delta r$. In metallic systems, that ratio is expected to be $\sim 1$ in 
accordance with the Wiedemann-Franz (WF) law, and this is indeed seen for the compressed HMJ, where the molecule is "metallic". However, upon stretching the molecule there is a significant deviation 
from the WF law. Thus, the dependence of $ZT$ on $\Delta r$, plotted in Fig.~\ref{fig7}(d), can be summarized as follows: the peak near $\Delta r\sim 0$\AA ~ is due to the enhanced slope of the transmission 
function caused by the shift of the LUMO resonance toward the Fermi level. The peak at large $\Delta r$ values is due both to the increase in $S$ (originating in the decrease in $\Gamma$) and to the 
violation of the WF law.

\section{Conclusion}
In Summary, we have shown that helicene-based molecular junctions can be used as a benchmark tool to investigate and mechanically tune transport and thermopower in molecular junctions. Using 
a combination of DFT and tight-binding calculations, we have shown that by compressing or stretching the molecular junction, its electronic properties can be controlled, leading to drastic 
changes in the transport (from a so-called "insulating" to "metallic" resonant-tunneling 
behavior) and possible substantial increase in the thermopower and thermoelectric efficiency. The origin of these effects is the helical spring-like structure of the helicene molecule; due to this 
structure, stretching and compressing leads to changes in the tunneling matrix elements between vertically-neighboring atoms (Fig.~\ref{fig1}). Control over the helicene 
length and number of rings was shown to lead to more than an order of magnitude increase in the thermoelectric figure of merit of HMJs, and to a typical thermopower which is an order of 
magnitude larger than that of more common molecular junctions. The idea of a mechanically controllable conductance and thermopower in molecular junctions should apply for other 
molecules with non-planar geometry, for instance cycloparaphenylenes \cite{Jasti2008,Sisto2011,Xia2012}, cyclacenes \cite{Choi1999}, ball-like molecules \cite{Kayahara2013,Kong2009}, 
carbon cages \cite{Wang2014,Matsui2014,Iwamoto2011}, tailored Fullerens \cite{Shustova2011}, Fullerene cages \cite{Skwara2011,Pomogaev2014,Hu2005}, and short DNA molecules (already demonstrated as elements in molecular junctions \cite{Wang2014a,Kang2010}, paving the way 
towards {\sl in-situ} mechanical control of transport properties of molecular junctions.  

\vskip 1truecm
The authors thank Prof. P. Reddy from the University of Michigan for valuable discussions and comments. 
Y.D. acknowledge funding from the University of Michigan -- Ben-Gurion University of the Negev Joint Research Collaboration. This work was financially supported by a Czech Science Foundation grant (P207/10/2207) and by the Institute of Organic Chemistry and Biochemistry, Academy of Sciences of the Czech Republic (RVO: 61388963).


%

\end{document}